\documentstyle[epsfig]{aipproc}

\begin{document}
\title{Physics in 2006}

\author{S. Dawson}
\address{Physics Department\thanks{Supported by the U.S.
Department of Energy under Contract No.
DE-AC02-76CH00016.}\\
Brookhaven National Laboratory\\
Upton, N.Y.~~ 11973}
\maketitle

\begin{abstract}
			Any consideration of future physics facilities must be 
made in the context of the Tevatron and the LHC.  I discuss some examples of 
physics results which could emerge from these machines and the resulting 
questions which would remain for a high energy $e^+e^-$ collider.  Particular
attention is paid to the electroweak symmetry breaking sector.  If
a light Higgs boson exists, it will be observed at the LHC and the 
role of any later accelerator will be to map out the Higg's
boson mass and couplings
and then
determine the space of possible models.  If there is no light Higgs
boson
then some effects of a strongly interacting electroweak symmetry breaking
sector will be observed at the LHC and I discuss 
the role of  a high energy
linear collider in exploring this scenario. 

\end{abstract}

\section*{Introduction}
Physicists are presently faced with a quandry.  Plans
and designs for the next generation of accelerators need to be formulated
in the near future, since the construction of these
machines spans many years.  The ``best'' machine
to build for the post-LHC era will only become clear, however,
 when we see what surprises the LHC holds. 
Faced with our imperfect knowledge, we must examine possible scenarios for 
LHC physics and determine the machine most likely to answer the
physics questions remaining in the years following completion of the LHC
and the fulfillment of its physics promise.  

Here, I will consider the physics of electroweak symmetry breaking at a high 
energy $e^+e^-$ collider.\footnote{For values of the Higgs boson
mass near
and slightly above 
$100~GeV$, a $\mu^+\mu^-$ collider operating at the Higgs resonance 
can make extremely precise measurements of the Higgs mass and couplings.
I will not discuss the physics potential of a muon collider here.\cite{muon}} 
This discussion must be made in the
 context of potential discoveries at the Tevatron
and the  LHC.  I begin by reviewing the
current experimental status of electroweak symmetry breaking,
 both from direct Higgs boson searches
 at LEP2 and from precision electroweak measurements.
Theoretical expectations for the Higgs boson mass are then reviewed,
 with emphasis on the implications for physics at higher mass scales.

Next, I review the discovery prospects
for the Higgs boson at both the Tevatron and the LHC.  The working
hypothesis  is that a weakly interacting Higgs boson will be discovered,
if it exists, at the Tevatron or the LHC, 
and so the role of the next generation
of accelerators will be to study the properties of a Higgs boson.
Precision
measurements of the mass, decay widths, and production rates
will all be  necessary 
in order 
to verify that a particle is the Higgs boson of the Standard Model.

Aside from the couplings to fermions and gauge bosons, 
we would also like to know that the Higgs boson self-interactions
result from the spontaneously broken
scalar potential of the Standard Model.  In order
to do this, the three- and four-
point self-couplings of the Higgs boson must
be measured.
These couplings can only be probed by multi-Higgs production,
which has extremely small rates, both at the LHC and at a high
energy $e^+e^-$ collider. 

The focus in this note is on verifying 
the properties of the Standard Model Higgs boson.  In order to do this, it is
helpful to compare with the predictions of a supersymmetric model since
these predictions may be  
quite different from those of the Standard Model.
Distinguishing between the Standard Model and a supersymmetric model
is an important test of our understanding of the electroweak sector.

If  a Higgs boson is not found at the Tevatron or the LHC, the electroweak
symmetry breaking sector must be strongly interacting.
 I end with a brief discussion of strong
electroweak symmetry breaking and a view towards the future.  

\section*{Inferences from the Standard Model}

\begin{figure}[b!] 
\centerline{\epsfig{file=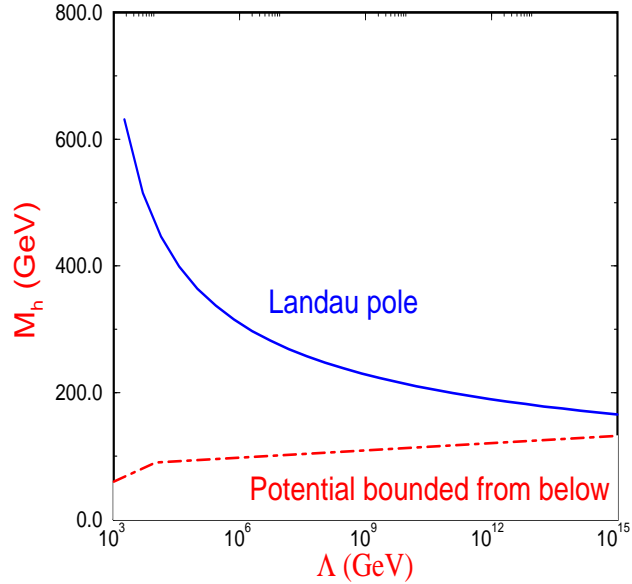,height=3.5in,width=3.5in}}
\caption{Theoretical expectations for a Standard Model Higgs boson,
as a function of the scale $\Lambda$ above which the Standard
Model is no longer valid. The region above the
solid curve has $\lambda(\Lambda)\rightarrow
\infty$, while the region below the dotted line has
$\lambda(\Lambda) < 0$.   The allowed region is the region 
between the curves.}
\label{fig1}
\end{figure}

\begin{figure}[b,t] 
\vskip -.5in
\centerline{\epsfig{file=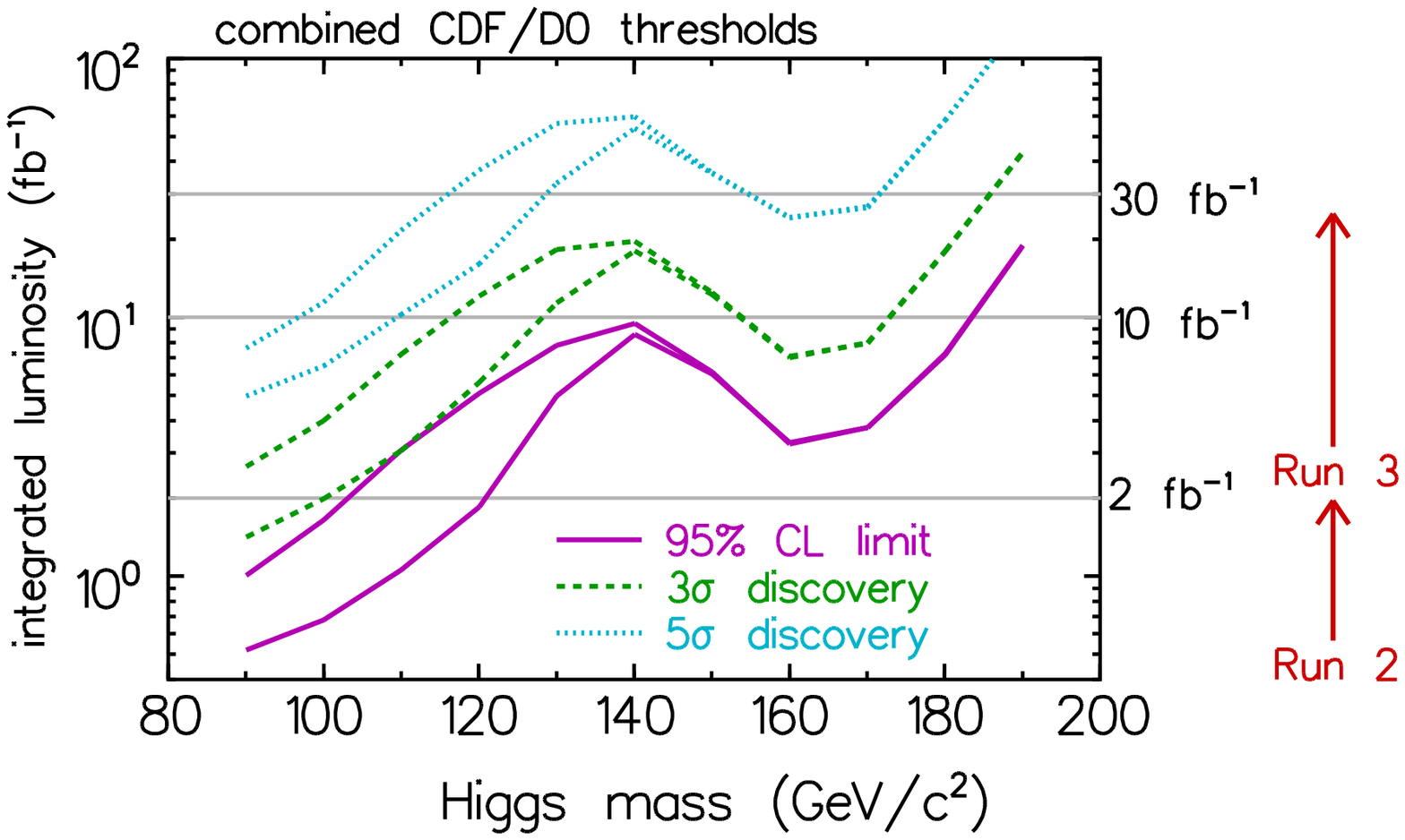,height=5.5in,width=4.5in}}
\vspace{10pt}
\vskip -1.5in
\caption{Discovery potential for a Standard Model Higgs boson at the
Tevatron. The vertical axis shows the required luminosity to
observe or exclude a given Higgs boson mass.  At the lower values
of the Higgs mass two curves are shown; the lower in each set is
a neural net analysis, the upper a standard analysis with cuts on
the signal and background. From Ref. 11.
}
\label{fig2}
\end{figure} 

The Standard Model  of electroweak interactions has been verified 
 to the $.1\%$ level through precision measurements
 at LEP and SLD.\cite{prerev}  In fact, the 
mechanism of electroweak symmetry breaking
 remains the only unconfirmed area of the 
Standard Model.  The Standard Model predicts
the existence of
 a physical scalar particle,
termed the Higgs boson.  
The search for this particle is therefore
a fundamental goal of all current and
future accelerators since its discovery is
needed to complete our knowledge of
the electroweak sector.
  The mass  is a 
free parameter of the theory and so the Higgs boson must be systematically
sought in all mass regions.

The couplings  of the scalar Higgs boson, however,
 are completely specified in terms of the
Higgs vacuum expectation value, $v=246~GeV$.
Hence branching ratios and production rates can be computed unambiguously in
terms of the  mass.
 Measurements of ratios of branching rates can then  be used
to test the validity of the model.  

Since the Higgs boson contributes to electroweak radiative corrections
at one loop,
precision measurements from LEP and SLD can be used to infer a prefered
value for the Higgs mass.  
The contribution of the Higgs boson to electroweak observables is
logarithmic and so the limit on the Higgs mass is not
nearly as precise as the indirect limit on the top quark mass
from precision measurements.
The current $95\%$  confidence level limit is,\cite{prerev} 
\begin{equation}
M_h < 230~GeV, ~~~~~{\hbox {Precision Measurements}}.
\end{equation}
It is important to understand that this limit assumes the validity of
the Standard Model.  Quantum loops containing new
particles can change this limit, as can new operators
beyond those of the Standard
Model. If there is new physics at the $TeV$ scale,
the limit of Eq. 1 can be evaded.\cite{alam}

The Higgs boson mass is the only free parameter of the 
electroweak theory.  
Although we cannot compute its mass, there are
certain theoretical restrictions following from the 
consistency of the theory.
  The scalar potential for an $SU(2)$ scalar doublet $\Phi$ is,
\begin{equation}
V=-\mu^2\mid \Phi\mid ^2+\lambda(\mid \Phi\mid^2)^2
\quad .
\label{potew}
\end{equation}
After the electroweak symmetry breaking has occured, there
remains the physical scalar Higgs boson $h$.
The quartic coupling, $\lambda$, is 
related to the Higgs boson mass, 
\begin{equation}
\lambda={M_h^2\over 2 v^2}\quad.
\end{equation}
Now $\lambda$ is not a fixed parameter, but 
scales with the relevent energy, $Q$, and  so
Eq. \ref{potew} is the
potential at the electroweak scale.
If $\lambda$ is large, (corresponding to a heavy Higgs boson), 
then at a scale $Q$,\cite{quiros}
\begin{equation}
Q{d \lambda\over d Q}=
{3\over 4\pi^2} \lambda,
\end{equation}
which can be solved to obtain, 
\begin{equation}
{1\over \lambda(\Lambda)}={1\over \lambda(M_h)
} 
-{3\over 4\pi^2}\log({\Lambda^2\over M_h^2})
\quad .
\end{equation}
A sensible theory will have $\lambda(\Lambda)$ finite at all scales, 
($\lambda(\Lambda)\rightarrow \infty$ is termed the Landau Pole),  
or correspondingly
${1\over \lambda(\Lambda)}>0$.  This yields an upper bound on $\lambda$
and hence on $M_h^2$,
\begin{equation}
M_h^2 < {8 \pi^2 v^2\over 3 \log(\Lambda^2/M_h^2)}\quad .
\end{equation}
If the Standard Model is valid to the GUT scale, $\Lambda\sim 10^{16}~GeV$,
then we have an approximate upper bound on the 
Higgs mass,\cite{quiros,chiv}
\begin{equation}
M_h < 170~GeV\quad .
\end{equation}
For any given value of $\Lambda$, there is a corresponding upper bound on 
$M_h$.  $\Lambda$ is often termed the ``scale of
new physics'' since above this scale, the Standard Model
interactions are not valid. This bound is the upper curve on Figure 1.

There is also a theoretical lower bound on $M_h$.  If $\lambda$
is small (light $M_h$), then 
\begin{equation}
Q{d\lambda\over dQ} \sim {1\over 16 \pi^2}
(B-12 g_t^4), 
\label{scalelh}
\end{equation}
where $g_t$ is the Higgs-top quark Yukawa coupling, ${g_t=-M_t/v}$,
and $B$ is a function of the gauge coupling constants, 
$~B={3\over 16}(2 g^4+( g^2+g^{\prime 2})^2)$. 
We see that the large top quark mass tends to drive $\lambda$ negative.  In
order for electroweak symmetry breaking to occur, the potential must
remain bounded from below, and $\lambda$ positive.   
Solving Eq. \ref{scalelh},
\begin{equation}
\lambda(\Lambda)=\lambda(M_h)+{B-12g_t^4\over 16\pi^2}\log({\Lambda
\over M_h})\quad .
\end{equation} 
Requiring $\lambda(\Lambda)>0$
gives the lower bound on $M_h$,
\begin{equation}
{M_h^2\over 2 v^2}> {B-12 g_t^4\over 16\pi^2}\log({\Lambda\over M_h})
\quad .
\end{equation} 
For large $M_t$, this relation changes sign and the two-loop
renormalization group corrections are important to obtain the
numerical bound.  Requiring that the Standard Model
be valid to the GUT scale,
$\Lambda=10^{16}~GeV$, gives the restriction on the
Higgs boson mass\cite{shere}
\begin{equation}
M_h>130~GeV\quad .
\label{lowerlim}
\end{equation}
This is shown as the lower curve in Figure 1.
There have been many theoretical improvements to
the naive bounds presented above, but the bottom
line is the same:   If the Standard Model is valid to the GUT scale, then
\begin{equation}
130~GeV < M_h < 180~GeV , \qquad  \Lambda \sim 10^{16}~GeV .
\label{rangmh}
\end{equation}
A Higgs boson outside this mass region would be a signal for 
new physics at the corresponding scale $\Lambda$. 
The mass region of Eq. \ref{rangmh}
 is particularly interesting since
it could potentially be probed at the Tevatron with an
upgraded luminosity.

There are also absolute bounds on the Higgs boson mass which are
independent of the scale, $\Lambda$.  Unitarity of the $WW$ elastic
scattering amplitudes requires $M_h < 800~GeV$, while lattice calculations
obtain a similar bound, $M_h < 700~GeV$.\cite{latt}  All of the theoretical
bounds of this section predict a Higgs boson comfortably within
the discovery range of the LHC and so if the Standard Model is
correct, a Higgs boson discovery should be just around the corner.  

\section*{Prospects for Discovery}

The current $95\%$ confidence
level limit on the Higgs boson mass from direct searches at
LEP2 
using data from $\sqrt{s}=189-202~GeV$
is\cite{blondel}
\begin{equation}
M_h > 106~GeV, \qquad LEP2.
\end{equation}
This limit is not expected to improve substantially
with further running
 at LEP2.

The minimal supersymmetric model has two neutral Higgs bosons, 
$h^{SUSY}$ and $H^{SUSY}$, a charged Higgs, $H^\pm$, and
a pseudoscalar, $A$.   The structure of the 
supersymmetric
potential dictates that at
lowest order all the couplings can be expressed in terms of
two parameters, which are typically taken to be the pseudoscalar
mass, $M_A$, and the ratio of Higgs vacuum expectation values,
$\tan\beta$.  All masses can then be expressed in terms of
these two parameters.\cite{susyrev}

The experimental limit on the Higgs  boson mass in a supersymmetric
theory typically depends on $\tan\beta$.
If we require that the limit be valid for all $\tan\beta$, there
is  a slightly lower $95\%$ confidence
level  limit  than for the Standard Model
Higgs boson,\cite{blondel}
\begin{equation}
M_h^{SUSY}>90~GeV\quad LEP2.
\label{susylim}
\end{equation}

The minimal supersymmetric theory has the remarkable feature that there is an
upper bound on the lightest Higgs boson resulting
from the structure of the scalar potential.  This bound
is roughly 
\begin{equation}
M_h^{SUSY} < 110-130~GeV,
\end{equation}
 where
the exact value depends on assumptions about the
parameters of the theory.\cite{carena}  This
is tantalizingly close to the experimental limit of Eq. \ref{susylim}.
 We see that there is no overlap
between the expected mass of the lightest  Higgs boson
of a supersymmetric model  and the
Standard Model Higgs boson when $\Lambda \sim M_{GUT}$.
Hence an observation of the Higgs boson with even an 
imprecise value for its mass will help to distinguish between
the Standard Model and its minimal supersymmetric extention.

A Standard Model Higgs boson should be discovered
at the Tevatron or the LHC.  Due to the small rate, the Higgs boson
will be extraordinarily difficult to observe at the Tevatron.
The signal with the best signature  is
associated production with a $W^\pm$.
For $M_h\sim 120~GeV$, the cross section at $\sqrt{s}=2~TeV$ is 
$\sigma(p {\overline p}\rightarrow W^\pm h)\sim
.3~pb.$  Even 
with $10~fb^{-1}$, the $5\sigma$ discovery level is only $M_h\sim 100~GeV$,
below the current LEP2 limit.
This underscores the need for the highest possible luminosity.

Figure 2 illustrates the discovery potential for a Standard Model Higgs
boson at the Tevatron.\cite{hobbs}
 For $M_h < 140~GeV$, the dominant signal 
results from $p {\overline p}\rightarrow Wh, h\rightarrow b {\overline
b}$, while at higher Higgs masses, the decay $h\rightarrow W W^*$
becomes the most important.  The discovery reach plot combines
small signals from many different
channels.   In fact, the maximum $S/\sqrt{B}$
in any channel is $.9$ for ${\cal L}=1~ fb^{-1}$.  A Standard
Model Higgs 
discovery at the Tevatron will almost certainly require the full
$25-30~fb^{-1}$ of  upgraded luminosity.  

The LHC, on the other hand, should discover a Standard Model Higgs boson
in any mass region below $1~TeV$, as illustrated in Figure 3, even with
only $30~fb^{-1}$.\cite{atlastdr}
From $M_h\sim 120~GeV$ all the way up to $M_h\sim 700$ GeV, the
Higgs boson can be observed through the decay $h\rightarrow ZZ
\rightarrow 4l$.  The discovery reach can be extended up to
$M_h\sim 1~TeV$ through the channels $h\rightarrow ZZ\rightarrow
l^+l^- \nu {\overline \nu}$ and $h\rightarrow W^+W^-
\rightarrow l\nu~jet~jet$.
  With the full luminosity of $100~fb^{-1}$, the LHC
will see a Higgs signal in multiple channels for all possible masses.
The observation in multiple channels will allow preliminary measurements
of the Higgs coupling constants, as discussed in the next section.   

The Standard Model  points
to a Higgs boson in the $100-200~GeV$ mass range, while
its minimal supersymmetric extention suggests that the lightest
Higgs boson is just above the current experimental
limit.  In either case, such a light Higgs boson would
be kinematically accessible through the process $e^+e^- \rightarrow
h Z$ at an $e^+e^-$ collider with $\sqrt{s}\sim 350-500~GeV$.
The rates for Higgs production at an $e^+e^-$ collider are shown 
in Fig. \ref{figee}.  
For an $e^+e^-$ collider with $\sqrt{s}\sim 500~GeV$,
the dominant production mechanism is $e^+e^-\rightarrow Zh$ for
$M_h\sim 200~GeV$.
At higher energy, 
say $\sqrt{s}\sim 1~TeV$, the largest rate is from $e^+e^-\rightarrow
\nu {\overline \nu} h$
  In
the next section, we examine the capabilities
and the required luminosities for  linear colliders 
to measure the Higgs properties  and 
contrast these potential future measurements with what we will know
from the LHC.    

\begin{figure}[b!] 
\centerline{\epsfig{file=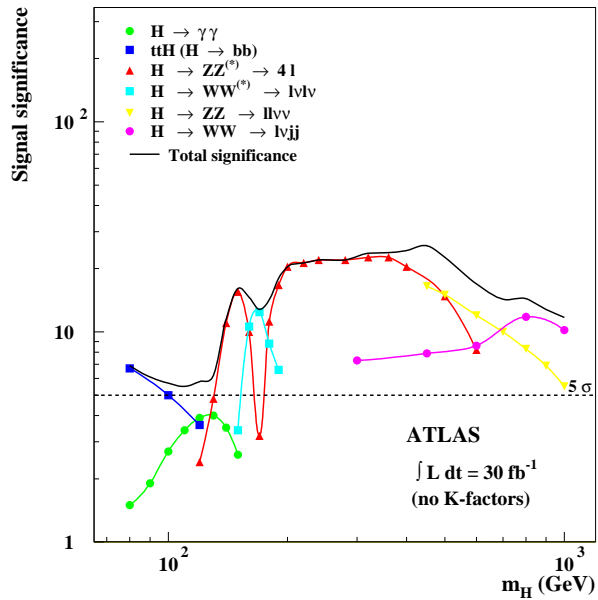,height=3.5in,width=3.5in}}
\vspace{10pt}
\caption{Discovery potential for a Standard Model Higgs boson at the
LHC, using the ATLAS detector, with $30~fb^{-1}$.
From Ref. 12.  }
\label{fig3}
\end{figure}

\begin{figure}[b,t] 
\centerline{\epsfig{file=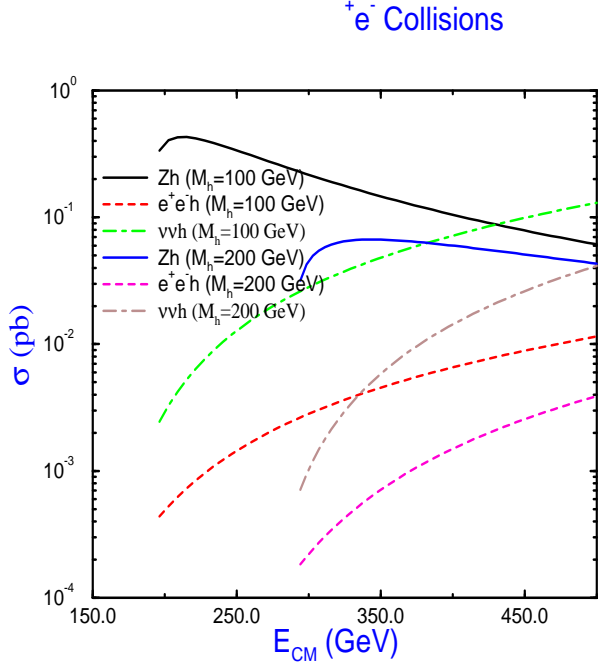,height=3.5in,width=3.5in}}
\vspace{10pt}
\caption{ Higgs boson  production at an  $e^+e^-$ collider.}
\label{figee}
\end{figure}

\section*{Precision measurements of mass, couplings, and branching ratios} 
\subsection*{Higgs Mass Measurements}

\begin{table}
\caption{Indirect Measurements of $M_h$}
\label{table1}
\begin{tabular}{lccc}
Collider  & $\Delta M_W$ & $\Delta M_t$ & ${\delta M_H\over M_H}$ \\ 
\tableline
  LEPII, TeV &  $30~MeV$& $4~GeV$ & $57~\%$\\ 
 LHC &  $15~MeV$ & $2~GeV$ & $26~\%$\\ 
 $500~GeV$ $e^+e^-$ &  $15~MeV$ & $200~MeV$ & $17~\%$\\ 
\tableline
\end{tabular}
\end{table}

\begin{figure}[b,t] 
\centerline{\epsfig{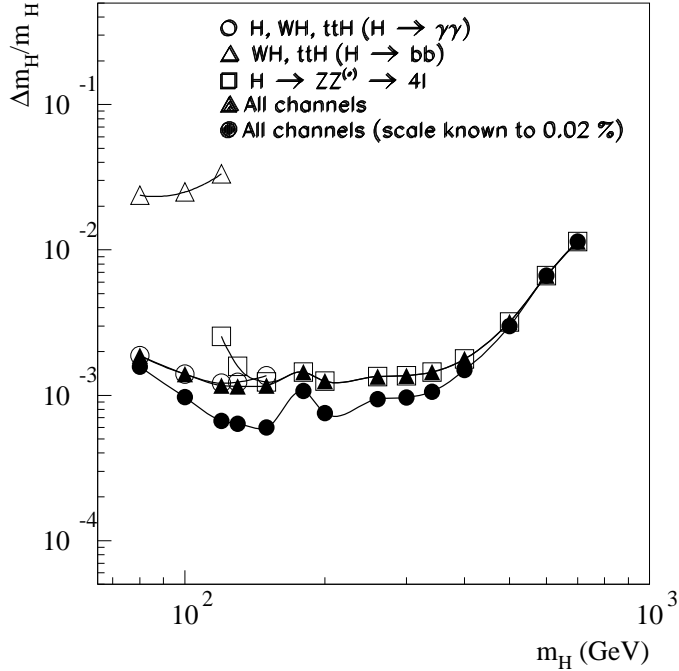}}
\vspace{10pt}
\caption{Precision
measurements of the Higgs
boson mass at the LHC with $\int {\cal L}=300~fb^{-1}$, using
the ATLAS detector.  From Ref. 12.}
\label{fig4}
\end{figure}

There are two complementary approaches to measuring the Higgs
boson mass.  The first is through
the direct observation of the Higgs boson.
 For most values of $M_h$, with an integrated  luminosity of $\int
{\cal L}=300~fb^{-1}$, the LHC will measure ${\delta M_h\over
M_h} \sim 10^{-3}$,
as shown in Fig. \ref{fig4}. 
Even at $M_h\sim 800~GeV$, the expected precision is ${\delta M_h
\over M_h} \sim
10^{-2}$.

At a high energy $e^+e^-$ collider, the cross section for $e^+e^-
\rightarrow Zh$ is a sensitive function of the Higgs boson mass and
we could hope to obtain an 
extremely precise measurement of the mass.  By
measuring the rate as a function of $\sqrt{s}$, a measurement 
of order \cite{tesla}
\begin{equation}
\delta M_h\sim 60~MeV \sqrt{{{\cal L}\over 100~fb^{-1}}}
\end{equation}
 could be obtained  for a
Higgs boson in the $100~GeV$ region.  
An alternate method is to
measure the recoil spectrum in the process
$e^+e^-\rightarrow Zh\rightarrow he^+e^-, h \mu^+\mu^-$.
This would yield
a precision of,
\begin{equation}
\delta M_h\sim 300~MeV \sqrt{{{\cal L}\over 100~fb^{-1}}}
\quad ,
\end{equation}
again for a Higgs boson in the $100~GeV$ region.
With $1000~fb^{-1}$ the precision on $\delta M_h$
for a light Higgs boson 
 at an $e^+e^-$ collider could
be considerably better than at the LHC, using either the excitation
spectrum or the recoil spectrum of the $e^+e^-\rightarrow Zh$ process.

Precise measurements of $M_W$ and $M_t$ 
at future colliders will allow a value of $M_h$ to be inferred\cite{tesla}, as
shown in Table 1.  (The $e^+e^-$ numbers in this table assume
$\int {\cal L}=1000~fb^{-1}$.)  Since the Higgs boson contributes
only logarithmically to electroweak  observables, the
precision is significantly less than the direct measurement.
Consistency between the direct and the indirect measurements
will
provide an important check of the theory at the quantum level,
however.

\subsection*{Measurements of Higgs Couplings}

\begin{figure}[b,t] 
\centerline{\epsfig{file=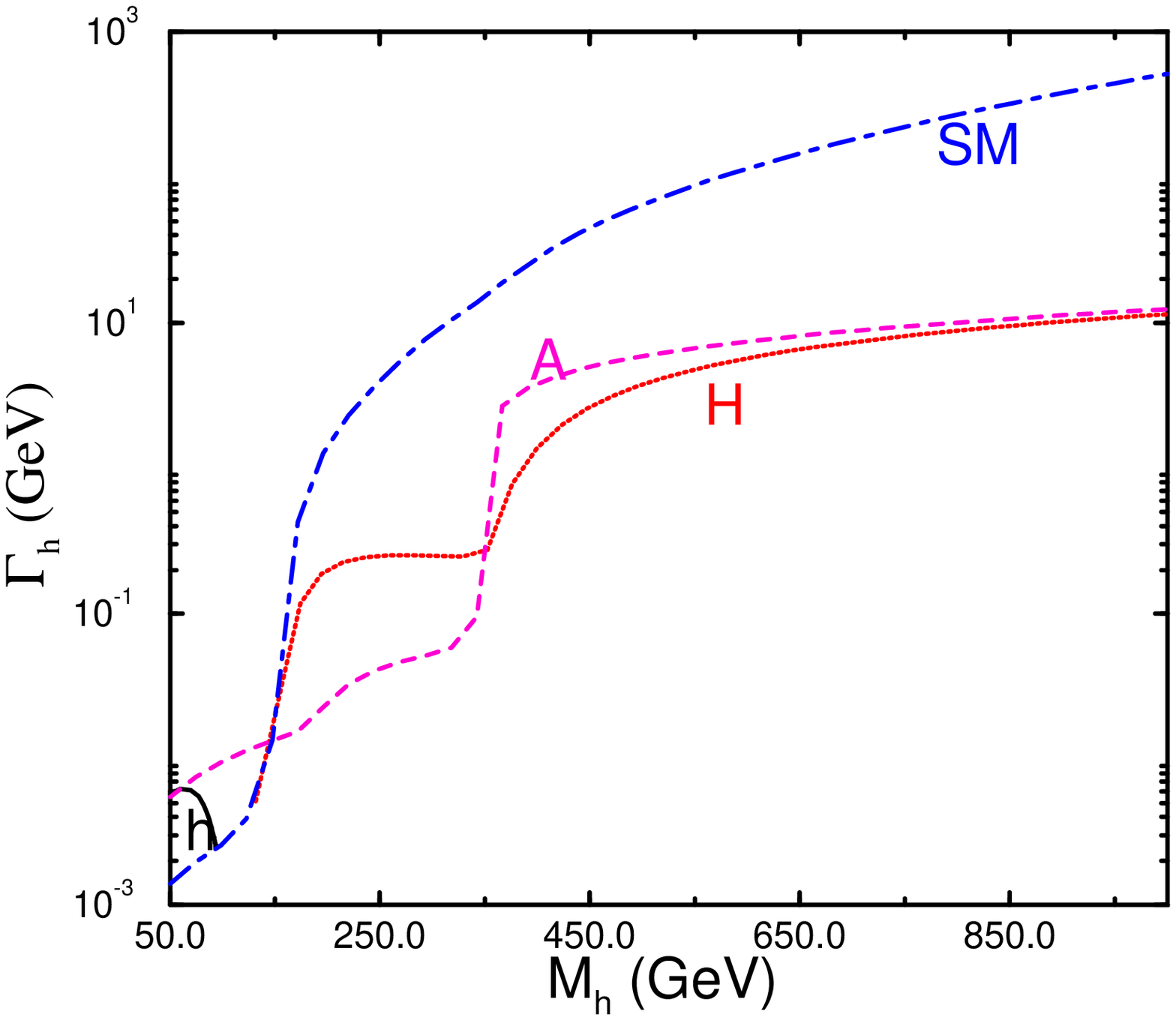,height=3.5in,width=3.5in}}
\vspace{10pt}
\caption{Total decay width for  Standard Model Higgs boson and
the Higgs bosons of a supersymmetric model
 with $\tan\beta=2$, $A_t=M_{SUSY}=1~TeV$,
and $\mu=100~GeV$.}
\label{fig5}
\end{figure}

The measurement of the Higgs boson couplings is important to differentiate 
between the Standard Model and other possibilities.  In a supersymmetric
model, the Higgs couplings to both fermions and gauge
bosons can be quite different from those of the Standard Model, as illustrated
in Fig. \ref{fig5} for an arbitrary choice of input parameters.  The
total decay width can differ by more than an order of magnitude 
between the Standard Model and a supersymmetric model. 

\begin{figure}[b,t] 
\centerline{\epsfig{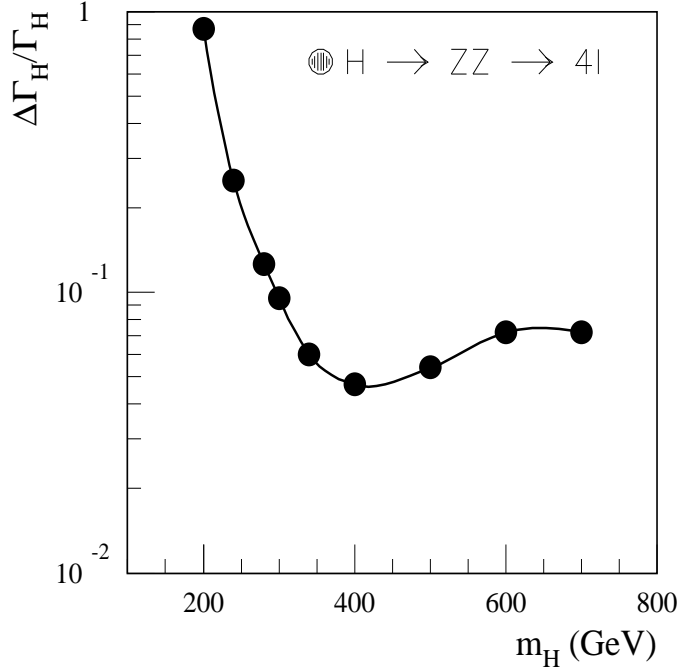}}
\vspace{10pt}
\caption{Measurement of the
total Higgs boson width at the LHC
with $300~fb^{-1}$, using the ATLAS detector.
From Ref. 12. }
\label{hwidfig}
\end{figure}

The total Higgs boson width can be measured from the reconstructed
Higgs peak at the LHC.  This direct measurement is only possible
for $M_h > 200~GeV$.  Below this mass, the width of the resonance
is narrower than the experimental resolution.  For $M_h > 200~GeV$,
the Higgs can be observed through the decay $h\rightarrow ZZ
\rightarrow 4l$ and the resulting measurement of the total width
is shown in Fig. \ref{hwidfig}.  With ${\cal L}=300~fb^{-1}$, the LHC
can measure $\Delta \Gamma_h/\Gamma_h < 10^{-1}$ for
$300~GeV < M_h < 800~GeV$.

\begin{figure}[b!] 
\centerline{\epsfig{file=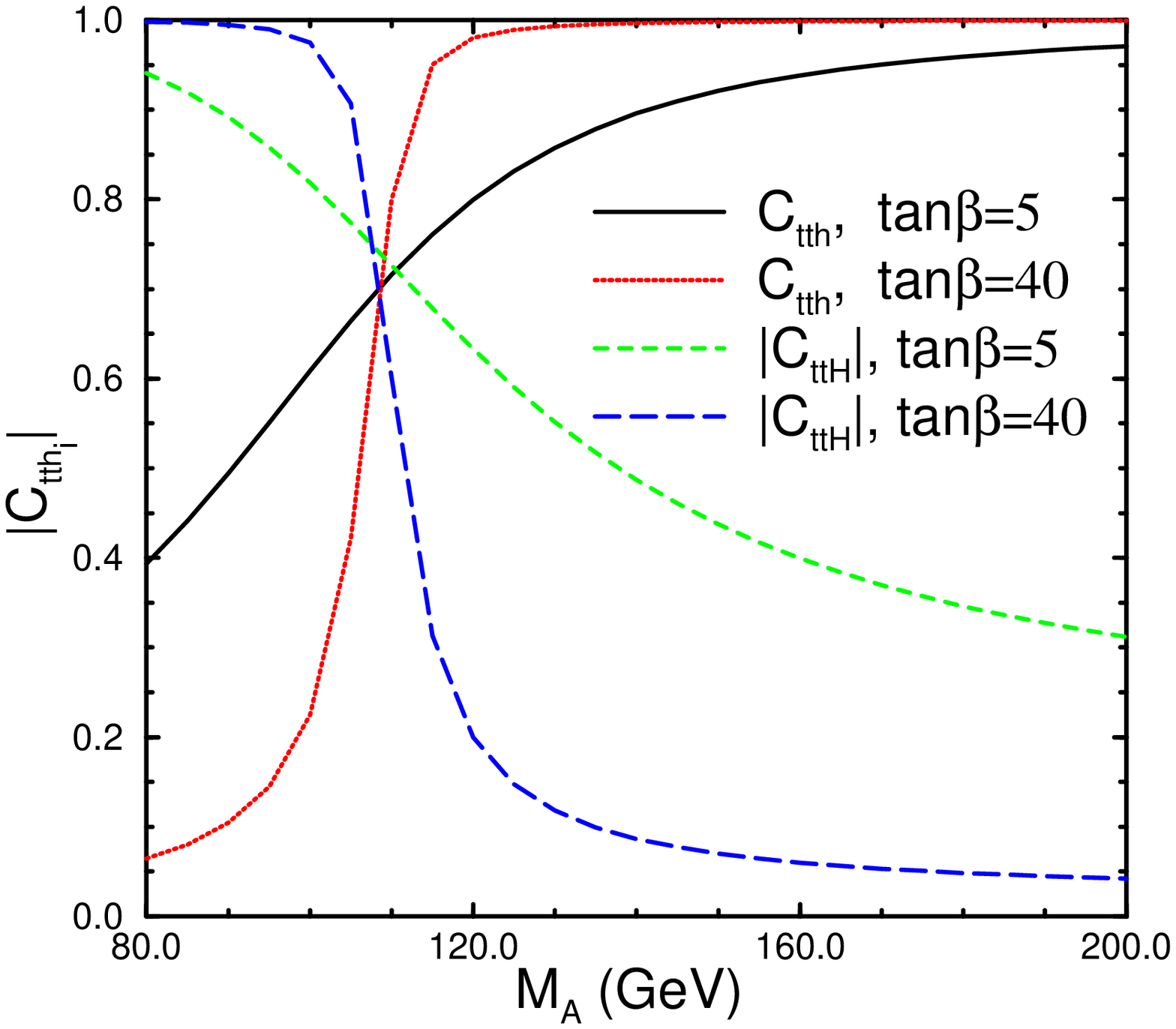,height=3.5in,width=3.5in}}
\vspace{10pt}
\caption{Couplings of the neutral Higgs bosons, $h^{SUSY}$
and  $H^{SUSY}$, of a
supersymmetric model to the top quark, in units of the Standard
Model Higgs-top quark Yukawa coupling.}
\label{fig6}
\end{figure}

Measurements of specific branching ratios are probably the most useful
quantities for distinguishing between the Standard Model and other
models.  
As an example, I discuss the coupling of the Higgs boson to the top
quark.  In the Standard Model the Yukawa coupling is given by,
\begin{equation}
g_t=-{M_t\over v},
\end{equation}
while in the minimal supersymmetric model the coupling is modified by the
factor $C_{tth}$,
\begin{equation}
g_t=-C_{tth}{M_t\over v} \quad .
\end{equation}
For some values of $\tan\beta$ and $M_A$, $C_{tth}$ can be quite different
from 1, as shown in Fig. \ref{fig6}.  Fig. \ref{fig6} also shows the
coupling of the heavier neutral Higgs boson of a supersymmetric
theory, $H^{SUSY}$,  to the top quark.  Again, the coupling can
be far from the Standard Model coupling.
Note that for $M_A\rightarrow \infty$, $C_{tth}\rightarrow 1$,
$C_{ttH}\rightarrow 0$  and
the Standard Model coupling is recovered.

At the LHC, the $t{\overline t}h$
 coupling can be measured to roughly $20\%$ through
the process $pp\rightarrow t {\overline t}h$ in the mass region $M_h\sim
120~GeV$.\cite{atlastdr} (For higher Higgs masses,
the cross section becomes quite small.)
 A similar mass region can be probed at an
$e^+e^-$ collider. The signal decays
predominantly to $W^+W^- b {\overline b} b {\overline b}$
and so will be  spectacular.  
A study of the signal and background showed that the signal could
be extracted from the background using both the semi-leptonic
and the hadronic decays of the $W$'s and a measurement of $g_t$  
obtained.\cite{morr,bdr}
Table 2 shows the expected precision
for the measurement of $g_t$
at $\sqrt{s}=500~GeV$ and $1~TeV$.\cite{bdr}  The message
is clear.  A precision measurement of $C_{tth} $ requires high energy and
high luminosity ($L=1000~fb^{-1}$) in order to improve on the LHC's
measurement.

\begin{table}
\caption{${\delta g_{t}\over g_t}$ in $e^+e^-$ interactions
With $1000~fb^{-1}$. From Ref. 15. }
\label{table2}
\begin{tabular}{lcc}
   $M_h~(GeV)$& $\sqrt{s}=500~GeV$ & $\sqrt{s}=1~TeV$ \\
\tableline
100	& .08	& .06 \\
110	& .12	& .06 \\
120	&.21	& .07 \\
130	& .44	& .08  \\
\tableline
\end{tabular}
\end{table}

The total rate for Higgs production in the process $e^+e^-\rightarrow
Zh$ can be found by measuring the recoil mass of the lepton pair,
$M_{ll}$, from the decay $Z\rightarrow l^+l^-$.  This measurement
is independent of the Higgs boson decay mode.  Once the total rate is
known, the Higgs branching ratios can be
measured by flavor tagging of the Higgs decay final states.

The measurements of the Higgs couplings to the lighter quarks can be 
done with a precision of $5-10~\%$ with $500~fb^{-1}$ at an $e^+e^-$
collider.  Ref. \cite{batt} found roughly equivalent results
for $\sqrt{s}=350~GeV$ and $\sqrt{s}=500~GeV$.  The error on the
measurements of the 
Higgs Yukawa couplings of Ref. \cite{batt} 
is dominated by theoretical uncertainty due to
the measured input values of $\alpha_s$,
$m_c$, and $m_b$, not by systematic or 
statistical errors.

\begin{figure}[b,t] 
\centerline{\epsfig{file=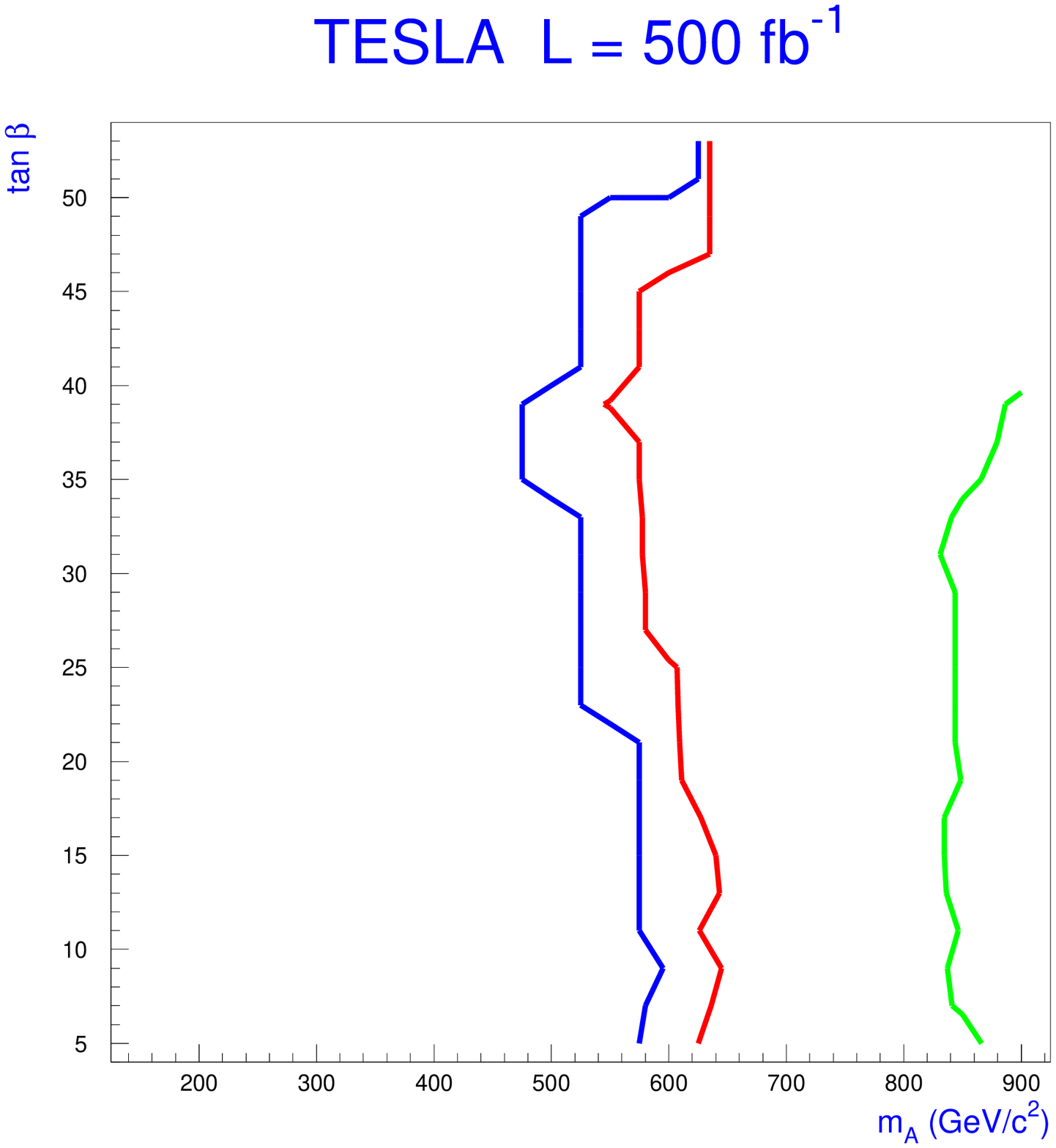,height=3.5in,width=3.5in}}
\vspace{10pt}
\caption{Regions  of  parameter space where  the Standard Model
 and the minimal supersymmetric model can
be distinguished. The regions to the right of
the curves (moving from left to right) have
more than $65\%$, $90\%$, or $95\%$ of the minimal
supersymmetric model solutions outside of the Standard
Model $95\%$  confidence level region.  From Ref. 16.} 
\label{fig8}
\end{figure}

Armed with measurements of the Higgs boson branching ratios, we can ask over what region of parameter
space the 
minimal supersymmetric model 
 can be distinguished from the Standard Model.
  The answer is shown
in Figure \ref{fig8}, taken from Ref. \cite{batt}.
First, the $95\%$ confidence level value of the branching ratio
for the Standard Model was computed.  Ref. \cite{batt} then
scans  over the parameter space of the minimal supersymmetric
model, taking $\tan\beta < 60$ and the mass parameters 
 to be
less than $1-1.5~TeV$.  For a given set of parameters, the Higgs
branching ratio was then computed.  In Fig. \ref{fig8}, the
region to the right of the curves (going from left to right on
the figure) has more than $68$, $90$ or $95\%$ of the supersymmetric
model solutions outside of the Standard Model $95\%$ confidence
level region. 
 With $500~fb^{-1}$, an $e^+e^-$ collider
can distinguish between the Standard Model
 and the minimal supersymmetric model  up to $M_A\sim 550~GeV$,
while with $1000~fb^{-1}$, 
 the sensitivity is increased to $M_A\sim 730~GeV$.\cite{batt}
This is remarkable given the decoupling of the Higgs sector of 
the minimal supersymmetric model for large $M_A$.

At the LHC, measurements of the Higgs couplings  are less clearcut
than at an $e^+e^-$ collider.  At the LHC, measurements involving the
Higgs boson typically involve combinations of Higgs couplings.  
For example, a measurement of the ratio of the $h\rightarrow \gamma\gamma$ and
$h\rightarrow ZZ\rightarrow 4l$ rates would give the ratio of the $h
\rightarrow \gamma\gamma$ and $h\rightarrow ZZ$ branching ratios, but not
the absolute couplings.  A study of the combinations of Higgs couplings
which can be measured at the LHC is given in Ref. \cite{snow1}.  

\section*{Verifying the structure of the Higgs potential}
\begin{figure}[b,t] 
\centerline{\epsfig{file=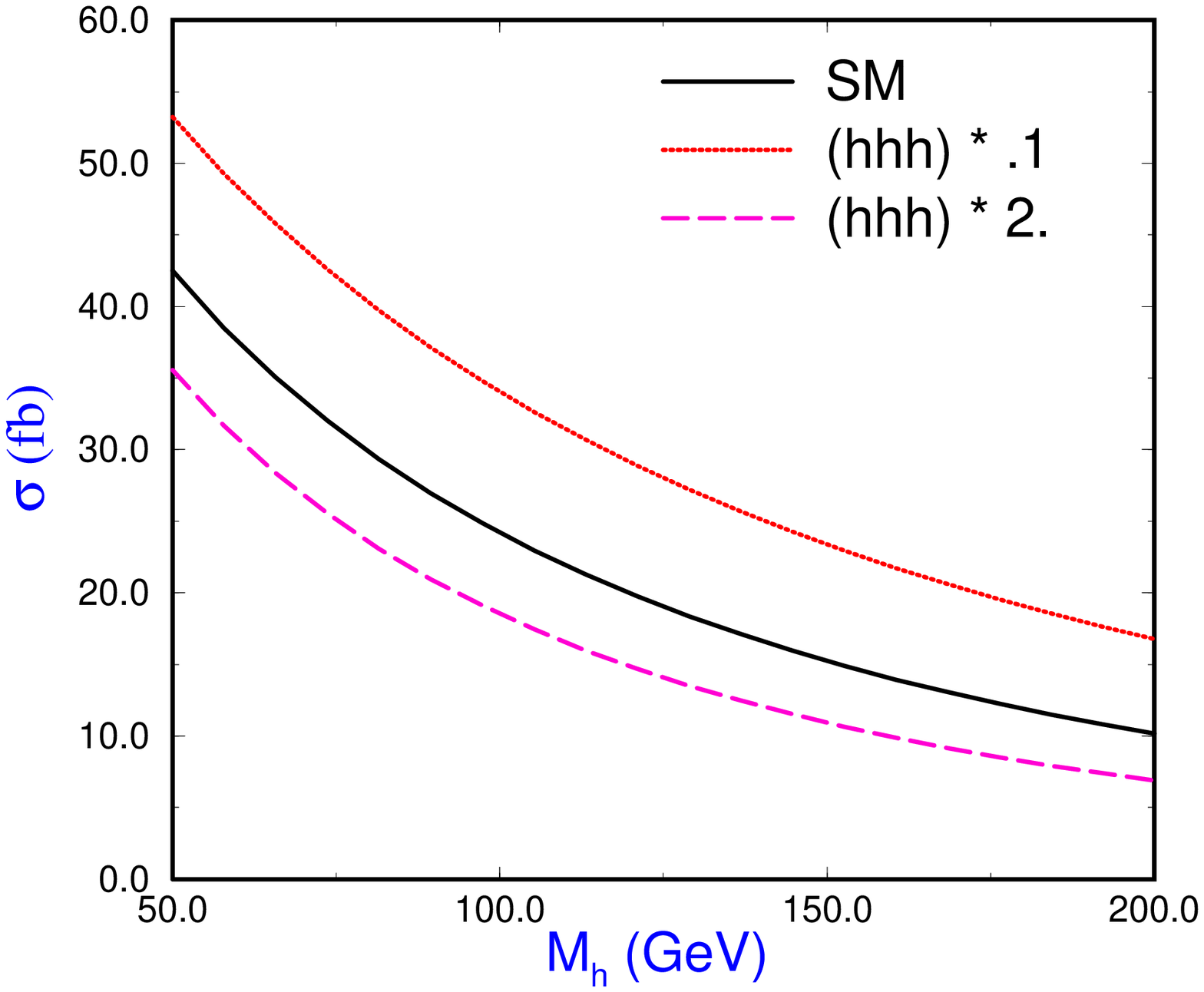,height=3.5in,width=3.5in}}
\vspace{10pt}
\caption{
Double Higgs production, $pp\rightarrow hh$,  at the LHC, $\sqrt{s}=14~TeV$.
The solid line is the Standard Model rate, while the dotted 
and dashed lines have the tri-linear Higgs couplings modified.}
\label{fig9}
\end{figure}

\begin{figure}[b,t] 
\centerline{\epsfig{file=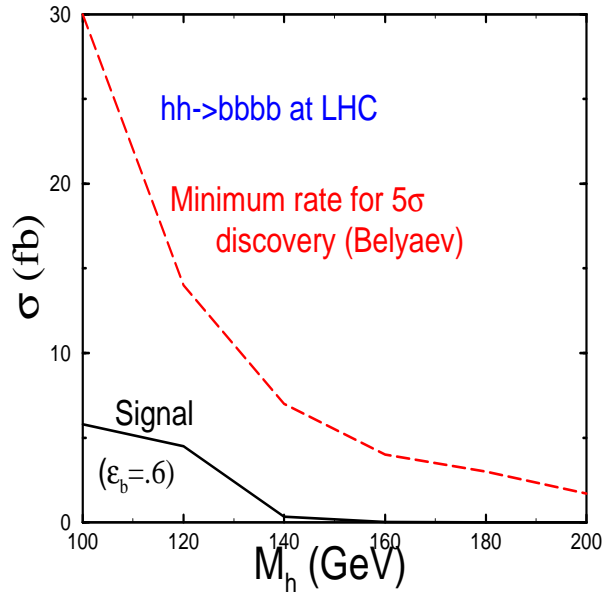,height=3.5in,width=3.5in}}
\vspace{10pt}
\caption{Minimum rate for a $5~\sigma$ discovery of $pp\rightarrow hhX$
at the LHC (dotted line) and the signal using a $b$ tagging efficiency
of $\epsilon_b=.6$.}
\label{fig10}
\end{figure}

\begin{figure}[b,t] 
\centerline{\epsfig{file=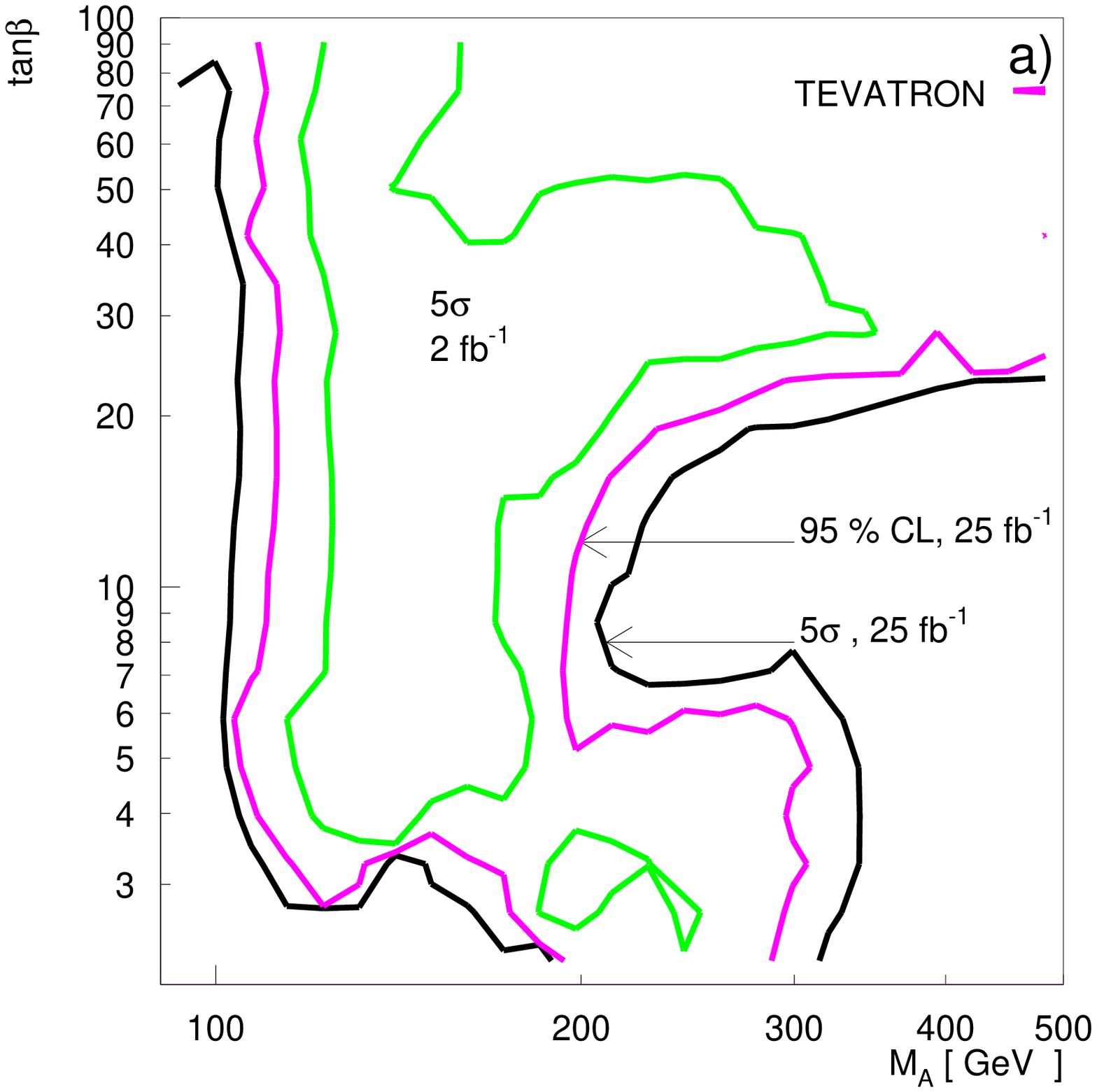,height=3.5in,width=3.5in}}
\vspace{10pt}
\caption{Double Higgs production  at the Tevatron
in a supersymmetric model 
 with enhanced squark contributions. From Ref. 18.}
\label{fighhtev}
\end{figure}

Once a Higgs particle is found,
it will be necessary
to investigate its self-couplings in order to reconstruct the Higgs potential
and 
to verify that the observed particle  is indeed the 
Standard Model
 Higgs boson which results from spontaneous symmetry breaking. A
first step in  this direction is the measurement of the trilinear
self-couplings of the Higgs boson
 which are uniquely specified by the scalar potential of Eq. 2.

After the symmetry breaking, the self-couplings of the Higgs boson
are uniquely determined by $M_h$, 
\begin{equation}
V={M_h^2\over 2}h^2 +{M_h^2\over 2 v} h^3+{M_h^2\over 
8 v^2} h^4\quad .
\end{equation}
 In extensions of the Standard Model, 
such as models with an extended scalar sector, with composite particles 
or with supersymmetric partners,
the self-couplings of the Higgs boson may be significantly different from
the Standard Model predictions. 

In order to probe the three- and four-
 point Higgs couplings, it is necessary
to measure multi-Higgs production.
Higgs boson 
pairs can be produced by several mechanisms at hadron colliders:
\begin{itemize}  
\item 
 Higgs-strahlung $W^*/Z^* \to h h W/Z$,
\item  
 vector-boson fusion $WW,ZZ \to hh$,
\item  
 Higgs radiation off top and bottom quarks $gg,q\bar q
 \to Q\bar Q h h $,
\item  
 gluon-gluon collisions $gg\to hh$.
\end{itemize}
At the LHC, gluon fusion
is the dominant source of 
Higgs-boson pairs in the Standard Model and arises from quark loops, with
the dominant contribution coming from top quark loops. The rate, 
even at the LHC, is quite
small as can be seen in Fig. 10.  
Although the rate is sensitive to the tri-linear coupling, the variation 
is probably too small to be observed.\cite{dds}
A detailed study of the signal and background
gives the results shown in Fig. 11.\cite{bely}
This study computed the minimum rate necessary for a $5\sigma$ discovery
of $hh$ production.
  It is clear that in the Standard Model,
this physics will have to wait for the next generation of accelerators.

In a supersymmetric
 model, the $b$- quark contribution to $hh$ production  will
be enhanced for large $\tan\beta$.  Even so, in the absence of large
squark loop contributions, with $25~fb^{-1}$ the Tevatron can only exclude
a small region of parameter space with $M_A<150~GeV$ and $\tan\beta > 80$.
The LHC will be able to exclude an even larger region of $M_A$ and
$\tan\beta$ space.\cite{bely} 
However, the situation changes dramatically for light squarks and with
the parameters chosen to maximize the squark tri-linear couplings.
In this case, it is possible to obtain a significant enhancement of
the rate, largely due to resonance effects.
This is shown in Fig. \ref{fighhtev}
 for the Tevatron.\cite{bely}
  In this very special situation,
even the Tevatron will be extremely sensitive to double Higgs production. 

At a high energy $e^+e^-$ collider, Higgs pairs are produced through
similar mechanisms as in hadronic collisions.
  At intermediate energies, $\sqrt{s} \sim 500~GeV$,
the dominant mechanism is $e^+e^-\rightarrow Zhh$, while at
TeV scale energies, the process $e^+e^-\rightarrow \nu {\overline \nu}
hh$ is dominant.  
Just above the kinematic  threshold, the sensitivity to the trilinear
coupling is maximal in the $e^+e^-\rightarrow Zhh$ process.  With
$2000~fb^{-1}$, the tri-linear coupling can be measured to 
$\sim 15\%$\cite{zerwas}.
The cross sections for
all sources of double Higgs production
in an $e^+e^-$ collider are small, on the order of a few
femptobarn or less
for $M_h < 200~GeV$.\cite{zerwas}
  This is clearly a measurement which requires the highest
possible luminosity in order to isolate the signal from
the background and make a measurement of the tri-linear
Higgs coupling.

At present, it does not appear possible to measure the
Higgs boson four-point coupling.  In principle, it could be measured in
triple Higgs  production, but the rate is miniscule.

\section*{Strongly Interacting Symmetry Breaking}

If a Higgs boson is not found at the LHC, then the electroweak symmetry
breaking is strongly interacting.  Without the addition of some new
type of physics, $WW$ scattering will violate unitarity
at an energy scale somewhere below $3~TeV$.  There are
two classes of effects which could potentially be observed in this
scenario.

The first possibility is that whatever new physics unitarizes the
$WW$ scattering is at too high an energy scale to be observed at
either the LHC or an $e^+e^-$ collider with $\sqrt{s}\sim 500~GeV-
1~TeV$.  In this case the only effects which can be observed are
small deviations in absolute rates.  The Lagrangian can be written
as 
\begin{equation}
{\cal L}={\cal L}_{SM}+\sum_i {f_i\over \Lambda^2}{\cal O}_i,
\end{equation}
where ${\cal L}_{SM}$ is the Lagrangian of the Standard Model with the Higgs
boson removed.  Without the Higgs boson, 
the Lagrangian can be written in terms of an
expansion in powers of ${s\over \Lambda^2}$, where $\Lambda$ is the
scale of new
physics.  The $f_i$ are dimensionless coefficients of the new
operators, ${\cal O}_i$.
  A complete set of operators at order $s/\Lambda^2$
 can be found in Ref. \cite{ab}.
The goal of the LHC or a high energy $e^+e^-$ collider in this
scenario would be to measure the $f_i$ and attempt to distinguish between
models.  At the LHC, there will be a very small number of events\cite{atlastdr}
and it is doubtful if it will be possible to tell the difference between
the various possible models.
An $e^+e^-$ collider with $\sqrt{s}\sim 1.5~TeV$ could measure some
of the $f_i$ to ${\cal O}(10^{-3})$, but a complete set of measurements
will take still higher energy.

In the second case, the new physics which unitarizes the $WW$ scattering
amplitudes produces resonances which can be observed.  Numerous studies have
found that an $e^+e^-$ collider with $\sqrt{s}\sim 1.5~TeV$ has
roughly the same sensitivity to $TeV$ scale resonances 
as does the LHC.\cite{snow1}  Both machines will be sensitive to resonances
on the order of $1.5~TeV$.   

\section*{Conclusion}

Even after the LHC has successfully run for a few years, there
will still be unanswered physics questions.  If a weakly interacting
Higgs boson exists, either from a supersymmetric model or the 
Standard Model, it will be observed at the LHC.
The LHC will make preliminary measurements of the Higgs boson mass and
couplings, but a high energy $e^+e^- $ collider with high luminosity
will significantly improve on the precision.  Precise measurements of
the Higgs width are particularly important for differentiating between
models. Measurements  of double Higgs production and strong symmetry
breaking in particular will require the highest possible energy and
luminosity.

This note has considered only electroweak symmetry breaking.  There will of
course be many exciting questions to be
answered in other areas of particle physics such as supersymmetry, 
QCD, CP violation, etc. Interesting times await us!

\end{document}